\documentclass[a4paper]{article}

\usepackage{pxfonts}
\usepackage{longtable}
\usepackage{natbib}
\usepackage{amssymb}
\usepackage{bussproofs}
\usepackage{url}

\newtheorem{definition}{Definition}
\newtheorem{theorem}{Theorem}

\usepackage{microtype}

\begin{document} 

\title{Normalisation for Bilateral Classical Logic with some Philosophical Remarks}
\author{Nils K\"urbis}
\date{}
\maketitle

\begin{center}
Published in the \emph{Journal of Applied Logics} 8/2 (2021): 531-556\\
\url{https://collegepublications.co.uk/ifcolog/?00044}\bigskip
\end{center}

\maketitle

\begin{abstract} 
\noindent Bilateralists hold that the meanings of the connectives are determined by rules of inference for their use in deductive reasoning with asserted and denied formulas. This paper presents two bilateral connectives comparable to Prior's \emph{tonk}, for which, unlike for \emph{tonk}, there are reduction steps for the removal of maximal formulas arising from introducing and eliminating formulas with those connectives as main operators. Adding either of them to bilateral classical logic results in an incoherent system. One way around this problem is to count formulas as maximal that are the conclusion of reductio and major premise of an elimination rule and to require their removability from deductions. The main part of the paper consists in a proof of a normalisation theorem for bilateral logic. The closing sections address philosophical concerns whether the proof provides a satisfactory solution to the problem at hand and confronts bilateralists with the dilemma that a bilateral notion of stability sits uneasily with the core bilateral thesis. 
\end{abstract}

\section{Introduction}
It is a commonly held view that the meanings of the expressions of a language are determined by the use its speakers make of them.\footnote{I thank Julien Dutant, Dorothy Edgington, Keith Hossack, Guy Longworth and Mark Textor for discussions about assertion and denial and audiences in Lecce, \L\'od\'z and Stirling for their comments on presentations of this paper.} One way of giving substance to this view is to propose that that use can be systematised for the hypothetical project of constructing a theory of meaning for a language in terms of the conditions of the correct assertibility of sentences containing the expressions. This is the course taken by Dummett \citep{dummettmeaningI, dummettmeaningII}. \emph{Bilateralism}, by contrast, is the view that the meanings of expressions are determined not only in terms of the conditions for the correct assertibility of sentences containing them, but by these in tandem with the conditions for their correct deniability. The view was proposed in response to Dummett by Price, who `takes the fundamental notion for a recursive theory of sense to be not assertion conditions alone, but these in conjunction with rejection, or \emph{denial} conditions' \citep[162]{pricesense}. We may distinguish the two views by calling the former \emph{unilateralism}. 

Unilateralism and bilateralism provide alternative forms for a theory of meaning for an entire language, but I will here only consider their restrictions to the logical constants of propositional logic. In that region of language, Dummett's insights coupled with important contributions by Prawitz have lead to the development of \emph{proof-theoretic semantics}, an alternative to truth-theoretic semantics. Whereas in the latter the meanings of the logical constants are given in terms of their contributions to the truth conditions of sentences containing them, the principal tenet of proof-theoretic semantics is that their meanings are determined by the use of such sentences in deductive arguments. 

In this paper I will present a problem for bilateral proof-theoretic semantics in the form of bilateral connectives that are comparable to Prior's \emph{tonk}. But whereas \emph{tonk} can be excluded from unilateral logic on principled grounds that form part of the philosophical background of proof-theoretic semantics, the issue is more involved in the case of bilateralism. The main part of the paper contains a proof of a normalisation theorem for a system of bilateral classical logic. This provides a solution of sorts to the problem, but it also has certain philosophical drawbacks. In particular, the proof appeals to an unrestricted version of a bilateral principle of non-contradiction, while Rumfitt requires this principle to be restricted to atomic premises. Secondly, the solution is based on a redefinition of the notion of a maximal formula, and it may be objected that the solution therefore merely constitutes a change of subject. I conclude that it would appear that the best solution appeals to bilateral analogues of Prawitz's inversion principle. These are desirable in any case and for independent reasons. Appeal to such principles, however, endangers the core thesis of bilateralism and threatens collapse it into unilateralism.

\section{A System of Bilateral Classical Logic} 
Proof-theoretic semantics along Dummett's and Prawitz's lines arguably does not go any further than intuitionist logic. From their perspective, the rules governing classical negation are defective. Advocates of bilateralism claim that this situation is rectified in their framework. They recommend the use of systems of natural deduction with two kinds of rules: For each connective $\mathbf{c}$, there are assertive rules specifying the grounds for and consequences of asserting a formula with $\mathbf{c}$ as main operator, and rejective rules specifying the grounds for and consequences of denying such a formula. The most prominent such system has been proposed by Rumfitt \citep{rumfittyesno}, building on work by Smiley \citep{smileyrejection}.\footnote{Humberstone proposed a similar system at the same time as Rumfitt \citep{humberstonerejection}.} Rumfitt's system is intended to satisfy Dummett's requirements for when the rules of inference governing a connective specify its meaning: they do so if they are \emph{in harmony} or, more precisely, \emph{stable} \citep[Chapters 11-13]{dummettLBM}. The aim is to provide `a direct specification of the senses of the connectives in terms of their deductive use' \citep[805]{rumfittyesno}, where the premises and conclusions of rules of inference are assertions and denials. 

Formulas in the system $\mathfrak{B}$ of bilateral classical logic are \emph{signed} by $+$, indicating asserted formulas, or $-$, indicating denied ones. $\bot$ indicates the incoherence that arises from asserting and denying the same formula. Deductions do not begin with $\bot$. Lower case Greek letters $\alpha, \beta$ range over signed formulas, $\phi$ may also be $\bot$. $\alpha^\ast$ designates the result of reversing $\alpha$'s sign from $+$ to $-$ or conversely. The terminology follows Rumfitt, and the rules of $\mathfrak{B}$ are his \citep[800ff]{rumfittyesno}. 

Deductions in $\mathfrak{B}$ have the familiar tree shape, with the (discharged or undischarged) assumptions at the top-most nodes or leaves and the conclusion at the bottom-most node or root. Every assumption in a deduction belongs to an \emph{assumption class}, marked by a natural number, different numbers for different assumption classes. Formula occurrences of different types must belong to different assumption classes. Formula occurrences of the same type may, but do not have to, belong to the same assumption class. Discharge of assumptions is marked by a square bracket around the formula: $[\alpha]^i$, where $i$ is a label for the assumption class to which $\alpha$ belongs. If the assumption is discharged, the label is repeated at the application of the rule. The formulas in an assumption class are discharged all together or not at all. Empty assumption classes are permitted for vacuous discharge, when a rule that allows for the discharge of assumptions is applied with no assumptions being discharged. The conclusion of a deduction is said to depend on the undischarged assumptions of the deduction. Similar terminology is applied to subdeductions of deductions. 

Upper case Greek letters $\Sigma$, $\Pi$, $\Xi$, possibly with subscripts or superscripts, denote deductions. Often some of the assumptions and the conclusion of the deduction are mentioned explicitly at the top and bottom of $\Sigma$, $\Pi$, $\Xi$. Using the same designation more than once to denote subdeductions of a deduction means that these subdeductions are exact duplicates of each other except that assumption classes may be different: the deductions have the same structure, and at every node formulas of the same type are premises and conclusions of applications of the same rules.\footnote{The layout of natural deduction used here follows \citep{troelstraschwichtenberg}.} 

\begin{definition}[Deduction in $\mathfrak{B}$]\ \\
\normalfont
\noindent (i) The formula occurrence $+ \ A\ ^n$ is a deduction in $\mathfrak{B}$ of $+ \ A$ from the undischarged assumption $+ \ A$, and $- \ A\ ^n$ is one of $- \ A$ from the undischarged assumption $- \ A$, where $n$ marks the assumption class to which $+\ A$, $- \ A$ belong.  

\noindent (ii) If $\Sigma$, $\Pi$, $\Xi$ are deductions in $\mathfrak{B}$, then so are the following, where the conclusion depends on the undischarged assumptions of $\Sigma$, $\Pi$, $\Xi$ except those in assumption classes $i$ and $j$: 

\noindent\begin{longtable}{l l} 
\bottomAlignProof\AxiomC{$\Pi$}\noLine\UnaryInfC{$+ \ A$}\AxiomC{$\Sigma$}\noLine\UnaryInfC{$+ \ B$}\LeftLabel{$+ \land I$:}\BinaryInfC{$+ \ A\land B$}\DisplayProof &\bottomAlignProof\AxiomC{$\Pi$}\noLine\UnaryInfC{$+ \ A\land B$}\LeftLabel{$+\land E$:}\UnaryInfC{$+ \ A$}\DisplayProof \quad \bottomAlignProof\AxiomC{$\Sigma$}\noLine\UnaryInfC{$+ \ A\land B$}\UnaryInfC{$+ \ B$}\DisplayProof \\
\\
\bottomAlignProof\AxiomC{$\Pi$}\noLine\UnaryInfC{$- \ A$}\LeftLabel{$-\land I$:}\UnaryInfC{$- \ A\land B$}\DisplayProof \quad \bottomAlignProof\AxiomC{$\Sigma$}\noLine\UnaryInfC{$- \ B$}\UnaryInfC{$- \ A\land B$}\DisplayProof  &  
 \bottomAlignProof\AxiomC{$\Xi$}\noLine\UnaryInfC{$- \ A\land B$} \AxiomC{$[- \ A]^i$}\noLine\UnaryInfC{$\Pi$}\noLine\UnaryInfC{$\phi$} \AxiomC{$[- \ B]^j$}\noLine\UnaryInfC{$\Sigma$}\noLine\UnaryInfC{$\phi$}\LeftLabel{$-\land E$:}\RightLabel{$_{i, j}$}\TrinaryInfC{$\phi$}\DisplayProof\\
\\
\bottomAlignProof\AxiomC{$\Pi$}\noLine\UnaryInfC{$+ \ A$}\LeftLabel{$+\lor I$:}\UnaryInfC{$+ \ A\lor B$}\DisplayProof \quad \bottomAlignProof\AxiomC{$\Sigma$}\noLine\UnaryInfC{$+ \ B$}\UnaryInfC{$+ \ A\lor B$}\DisplayProof & 
\bottomAlignProof\AxiomC{$\Xi$}\noLine\UnaryInfC{$+ \ A\lor B$} \AxiomC{$[+ \ A]^i$}\noLine\UnaryInfC{$\Pi$}\noLine\UnaryInfC{$\phi$} \AxiomC{$[+ \ B]^j$}\noLine\UnaryInfC{$\Sigma$}\noLine\UnaryInfC{$\phi$}\LeftLabel{$+\lor E$:}\RightLabel{$_{i, j}$}\TrinaryInfC{$\phi$}\DisplayProof\\
\\
\bottomAlignProof\AxiomC{$\Pi$}\noLine\UnaryInfC{$- \ A$}\AxiomC{$\Sigma$}\noLine\UnaryInfC{$- \ B$}\LeftLabel{$-\lor I$:}\BinaryInfC{$- \ A\lor B$}\DisplayProof  & \bottomAlignProof\AxiomC{$\Pi$}\noLine\UnaryInfC{$- \ A\lor B$}\LeftLabel{$-\lor E$:}\UnaryInfC{$- \ A$}\DisplayProof \quad \bottomAlignProof\AxiomC{$\Sigma$}\noLine\UnaryInfC{$- \ A\lor B$}\UnaryInfC{$- \ B$}\DisplayProof \\
\\
\bottomAlignProof\AxiomC{$[+ \ A]^i$}\noLine\UnaryInfC{$\Pi$}\noLine\UnaryInfC{$+ \ B$}\LeftLabel{$+\supset I$:}\RightLabel{$_i$}\UnaryInfC{$+ \ A\supset B$}\DisplayProof & \bottomAlignProof\AxiomC{$\Pi$}\noLine\UnaryInfC{$+ \ A\supset B$}\AxiomC{$\Sigma$}\noLine\UnaryInfC{$+ \ A$}\LeftLabel{$+\supset E$:}\BinaryInfC{$+ \ B$}\DisplayProof\\
\\
\bottomAlignProof\AxiomC{$\Pi$}\noLine\UnaryInfC{$+ \ A$}\AxiomC{$\Sigma$}\noLine\UnaryInfC{$- \ B$}\LeftLabel{$-\supset I$:}\BinaryInfC{$- \ A \supset B$}\DisplayProof &  \bottomAlignProof \AxiomC{$\Pi$}\noLine\UnaryInfC{$- \ A\supset B$}\LeftLabel{$-\supset E$:}\UnaryInfC{$+ \ A$}\DisplayProof \quad \bottomAlignProof\AxiomC{$\Sigma$}\noLine\UnaryInfC{$- \ A\supset B$}\UnaryInfC{$- \ B$}\DisplayProof\\
\\
\bottomAlignProof\AxiomC{$\Pi$}\noLine\UnaryInfC{$- \ A$}\LeftLabel{$+\neg I$:}\UnaryInfC{$+ \ \neg A$}\DisplayProof & \bottomAlignProof\AxiomC{$\Pi$}\noLine\UnaryInfC{$+ \ \neg A$}\LeftLabel{$+\neg E$:}\UnaryInfC{$- \ A$}\DisplayProof\\
\\
\bottomAlignProof\AxiomC{$\Pi$}\noLine\UnaryInfC{$+ \ A$}\LeftLabel{$-\neg I$:}\UnaryInfC{$- \ \neg A$}\DisplayProof & \bottomAlignProof\AxiomC{$\Pi$}\noLine\UnaryInfC{$- \ \neg A$}\LeftLabel{$-\neg E$:}\UnaryInfC{$+ \ A$}\DisplayProof\\
\\
\bottomAlignProof\AxiomC{$[\alpha^\ast]^i$}\noLine\UnaryInfC{$\Pi$}\noLine\UnaryInfC{$\bot$}\LeftLabel{Reductio: \ }\RightLabel{$_i$}\UnaryInfC{$\alpha$}\DisplayProof & \bottomAlignProof\AxiomC{$\Pi$}\noLine\UnaryInfC{$\alpha$}\AxiomC{$\Sigma$}\noLine\UnaryInfC{$\alpha\ast$}\LeftLabel{Non-Contradiction: \ }\BinaryInfC{$\bot$}\DisplayProof
\end{longtable}

\noindent (iii) Nothing else is a deduction in $\mathfrak{B}$. 
\end{definition} 

\noindent Rumfitt calls reductio and non-contradiction \emph{co-ordination principles}. They have the character of structural rules required by the formal framework of bilateral logic to regulate the interaction between $+$, $-$ and $\bot$.\footnote{An alternative version of reductio avoids the use of $\bot$: from $\Gamma, \alpha^\ast\vdash \beta$ and $\Delta, \alpha^\ast\vdash\beta^\ast$, infer $\Gamma, \Delta\vdash \alpha$. An intuitionist bilateral logic has been formalised in \citep{kurbisrumfitt}. It fulfils the requirements Rumfitt imposes on a satisfactory bilateral logic, and hence the claim that only classical bilateral logic can do so is false. For an informal argument against the view that bilateralism inevitably leads to classicism, see \citep{kurbisbilateraldetours}. The stipulation that nothing can be both asserted and denied addresses the problem with negation in intuitionist logic noted in \citep{kurbisnegation}.} 

According to Rumfitt, non-contradiction must be restricted to atomic premises \citep[815f]{rumfittyesno}. His reason is that on a bilateral account of meaning, only the atomic sentences are co-ordinated primitively by non-contradiction: it is a consequence of how their use is specified in terms of the conditions of their correct assertibility and deniability. That the complex sentences are also so co-ordinated is a consequence of co-ordination at the atomic level and how the meanings of the connectives are specified by their assertive and rejective rules. By contrast, I will not impose this restriction on non-contradiction. 

It is generally considered to be a necessary requirement for a system of natural deduction to be satisfactory from the perspective of proof-theoretic semantics that deductions in it \emph{normalise}. In unilateral logic, a deduction is in \emph{normal form} if it contains no \emph{maximal formulas}, where a maximal formula is one that is the conclusion of an introduction rule and major premise of an elimination rule for its main connective. This definition carries over to $\mathfrak{B}$, only that maximal formulas are signed by $+$ or $-$. In section 4 it will be shown that such formulas, and further undesirable ones, can indeed be removed from deductions in $\mathfrak{B}$. The proof of this more general result requires the unrestricted version of non-contradiction.

\section{Some Principles of Proof-Theoretic Semantics} 
Proof-theoretic semantics has its roots in a comment of Gentzen's, who formulated the rudiments of a theory of meaning for the connectives: 

\begin{quote}
The introductions constitute, so to speak, the ``definitions'' of the symbols concerned, and the eliminations are in the end only consequences thereof, which could be expressed thus: In the elimination of a symbol, the formula in question, whose outer symbol it concerns, may only ``be used as that which it means on the basis of the introduction of this symbol''. \citep[189]{gentzenuntersuchungen} 
\end{quote} 

\noindent Gentzen's comment is the foundation of Prawitz's \emph{inversion principle}: `an elimination rule is, in a sense, the inverse of the corresponding introduction rule: by an application of an elimination rule one essentially only restores what had already been established if the major premise of the application was inferred by an application of an introduction rule $[\ldots ;]$ nothing is ``gained'' by inferring a formula through introduction for use as a major premiss in an elimination.' \citep[33f]{prawitznaturaldeduction} Prawitz proposes the normalisability of deductions as a formal criterion for when the inversion principle is met. 

According to Dummett, the meanings of expressions are determined by two aspects of their use, their contributions to the grounds for asserting sentences in which they occur and to the consequences of asserting such sentences \citep[211ff]{dummettLBM}. The connectives are a particularly clear cases of how this insight may be applied. The introduction rules for a connective specify the \emph{canonical grounds} for deriving a formula with that connective as main operator, and its elimination rules specify the \emph{canonical consequences} that follow from such a formula. For the rules governing a connective to determine its meaning completely, the two aspects of their use must be stable. 

Prawitz's inversion principle captures the thought that the elimination rules for a connective $\mathbf{c}$ should not licence the deduction of more formulas from a formula with $\mathbf{c}$ as main operator than are justified by the grounds of its assertion as specified by its introduction rule. This is what's wrong with Prior's \emph{tonk} \citep{priorrunabout}: 

\begin{center}
\AxiomC{$A$}
\LeftLabel{\emph{tonk I}: \ } 
\UnaryInfC{$A \mathit{tonk} B$}
\DisplayProof\qquad\qquad
\AxiomC{$A \mathit{tonk} B$}
\LeftLabel{\emph{tonk E}: \ }
\UnaryInfC{$B$}
\DisplayProof
\end{center} 

\noindent The elimination rule of \emph{tonk} licences the derivation of too many consequences from $A \mathit{tonk} B$ relative to its introduction rule. Maximal formulas of the form $A\mathit{tonk} B$ cannot be removed: 

\begin{prooftree}
\AxiomC{$A$}
\UnaryInfC{$A\mathit{tonk} B$}
\UnaryInfC{$B$}
\end{prooftree}

\noindent The rules for \emph{tonk} do not satisfy Prawitz's inversion principle.  

Prawitz's inversion principle is not enough for meaning-theoretical purposes. Consider a connective with the introduction rule of conjunction but only one of its elimination rules. Something is missing: its elimination rule does not permit the use of the connective in all the ways one should be able to use it relative to its introduction rule. 

Prawitz's inversion principles spells out the notion of harmony. Dummett's notion of stability consists in harmony together with a suitable convers. The latter, as Moriconi and Tesconi note \citep[111]{moriconitesconi}, is provided by an inversion principle of Negri's and von Plato's: `Whatever follows from the direct grounds for deriving a proposition must follow from the proposition' \citep[6]{negriplatostructural}. The elimination rules for a connective should licence the deduction of all the consequences from a formula with that connective as main operator that are justified relative to its introduction rules.\footnote{For more on inversion principles, see \citep{milneinversion}.}  

Notice that \emph{tonk} satisfies Negri's and von Plato's inversion principle: whatever follows from the direct grounds for deriving $A\mathit{tonk} B$ follows from $A\mathit{tonk} B$. Consequently, as Prawitz's inversion principle is tied to normalisation, it is a notion interesting enough to be considered by itself, whereas a suitable converse of Prawitz's inversion principle, such as Negri's and von Plato's, is usually considered only in combination with Prawitz's. 

If both inversion principles are satisfied, stability obtains and the elimination rules for a connective licence the deduction of all and only the consequences from a sentence with the connective as main operator that are justified by the grounds for deriving it as specified by its introduction rules.

\section{Bilateral Dissonance} 
Consider the connective \emph{conk}:\bigskip 

\noindent\begin{tabular}{l l} 
$+ \mathit{conk} I$: \AxiomC{$+ \ A$}\AxiomC{$+ \ B$}\BinaryInfC{$+ \ A \mathit{conk} B$}\DisplayProof &
$+\mathit{conk} E$:  \AxiomC{$+ \ A\mathit{conk} B$}\UnaryInfC{$+ \ A$}\DisplayProof \ \AxiomC{$+ \ A\mathit{conk} B$}\UnaryInfC{$+ \ B$}\DisplayProof \\
\\
$-\mathit{conk}I$:  \AxiomC{$- \ A$}\AxiomC{$- \ B$}\BinaryInfC{$- \ A\mathit{conk} B$}\DisplayProof  & $-\mathit{conk} E$:  \AxiomC{$- \ A\mathit{conk} B$}\UnaryInfC{$- \ A$}\DisplayProof \ \AxiomC{$- \ A\mathit{conk} B$}\UnaryInfC{$- \ B$}\DisplayProof
\end{tabular}\bigskip

\noindent  \emph{conk} means trouble. Given reductio, the assertion of any formula follows from the assertion of any formula: 

\begin{prooftree}
\AxiomC{}
\RightLabel{$_1$}
\UnaryInfC{$- \ A\mathit{conk} B$}
\UnaryInfC{$- \ A$}
\AxiomC{$+ \ A$}
\BinaryInfC{$\bot$}
\UnaryInfC{$+ \ A\mathit{conk} B$}
\RightLabel{$_1$}
\UnaryInfC{$+ \ B$}
\end{prooftree} 

\noindent The denial of any formula also follows from the denial of any formula: 

\begin{prooftree}
\AxiomC{}
\RightLabel{$_1$}
\UnaryInfC{$+ \ A\mathit{conk} B$}
\UnaryInfC{$+ \ A$}
\AxiomC{$- \ A$}
\BinaryInfC{$\bot$}
\UnaryInfC{$- \ A\mathit{conk} B$}
\RightLabel{$_1$}
\UnaryInfC{$- \ B$}
\end{prooftree} 

\noindent Notice that \emph{conk} permits the restriction of non-contradiction to atomic premises. 

Next consider the connective \emph{honk}:\bigskip 

\noindent \begin{tabular}{l l} 
$+\mathit{honk}I$:  \AxiomC{$- \ A$}\AxiomC{$+ \ B$}\BinaryInfC{$+ \ A\mathit{honk} B$}\DisplayProof  & $+\mathit{honk} E$:  \AxiomC{$+ \ A\mathit{honk} B$}\UnaryInfC{$- \ A$}\DisplayProof \ \AxiomC{$+ \ A\mathit{honk} B$}\UnaryInfC{$+ \ B$}\DisplayProof \\
\\
$- \mathit{honk} I$: \AxiomC{$+ \ A$}\AxiomC{$- \ B$}\BinaryInfC{$- \ A \mathit{honk} B$}\DisplayProof &
$-\mathit{honk} E$:  \AxiomC{$- \ A\mathit{honk} B$}\UnaryInfC{$+ \ A$}\DisplayProof \ \AxiomC{$- \ A\mathit{honk} B$}\UnaryInfC{$- \ B$}\DisplayProof
\end{tabular}\bigskip

\noindent \emph{honk}, too, means trouble. Given reductio, the assertion of any formula follows from the denial of any formula: 

\begin{prooftree}
\AxiomC{}
\RightLabel{$_1$}
\UnaryInfC{$- \ A\mathit{honk} B$}
\UnaryInfC{$+ \ A$}
\AxiomC{$- \ A$}
\BinaryInfC{$\bot$}
\UnaryInfC{$+ \ A\mathit{honk} B$}
\RightLabel{$_1$}
\UnaryInfC{$+ \ B$}
\end{prooftree} 

\noindent The denial of any formula follows from the assertion of any formula: 

\begin{prooftree}
\AxiomC{}
\RightLabel{$_1$}
\UnaryInfC{$+ \ A\mathit{honk} B$}
\UnaryInfC{$- \ A$}
\AxiomC{$+ \ A$}
\BinaryInfC{$\bot$}
\UnaryInfC{$- \ A\mathit{honk} B$}
\RightLabel{$_1$}
\UnaryInfC{$- \ B$}
\end{prooftree} 

\noindent \emph{honk} also permits the restriction of non-contradiction to atomic premises. 

The rules for \emph{conk} and \emph{honk} appear to be just as good as the rules for the connectives of $\mathfrak{B}$. They combine old rules in novel ways. Like \emph{tonk}, \emph{conk} combines rules for conjunction and disjunction, only this time they are bilateral rules and all assertive rules for conjunction and all rejective rules for disjunction are used. \emph{honk} combines the rejective rules for implication with assertive rules that would be correct for another connective. But unlike \emph{tonk}, \emph{conk} and \emph{honk} have the rather unusual feature that although adding them to $\mathfrak{B}$ gives an incoherent system, maximal formulas that arise from concluding $A\mathit{conk}B$ or $A\mathit{honk}B$ by an introduction rule and using them as major premises of an elimination rule may be removed from deductions by the same reduction procedures that remove such maximal formulas with conjunction, disjunction or implication as main connectives. 

The rules for the connectives of $\mathfrak{B}$ satisfy bilateral versions of the inversion principles. The assertive elimination rules for a connective of $\mathfrak{B}$ licence the deduction of all and only the consequences from an asserted sentence with the connective as main operator that are justified by the grounds for deriving it as specified by its assertive introduction rules. The rejective elimination rules for a connective of $\mathfrak{B}$ licence the deduction of all and only the consequences from a denied sentence with the connective as main operator that are justified by the grounds for deriving it as specified by its rejective introduction rules. Unlike \emph{tonk}, the rules for \emph{conk} and \emph{honk} also satisfy these bilateral inversion principles.\footnote{Gabbay has also proposed a connective that satisfies the bilateral inversion principles but leads to incoherence \citep{gabbaybilateralism}. \emph{conk} and \emph{honk}, however, are worse than Gabbay's connective, as they satisfy an additional requirement concerning the proper subformulas of premises and conclusions of rules of inference. An exposition of the precise nature of this requirement and why it may reasonably be imposed on rules that are to determine the meanings of the connectives they govern completely is the subject of a piece currently in preparation.}

It is evident where the problem lies. In the bilateral framework, it is not enough that inversion principles balance the grounds and consequences of asserting a formula and others balance the grounds and consequences of denying a formula. There also needs to be a sort of stability between the assertive and the rejective rules for a connective, a kind of inversion that balances the grounds and consequences of the assertion and the denial of a formula.  

The issue can also be put in terms of the question why the rejective and the assertive rules for a connective of $\mathfrak{B}$ are rules for the same connective. What is it, for instance, that makes the assertive rules for the symbol $\land$ and the rejective rules for the symbol $\land$ rules for \emph{conjunction}? What justifies the use of the same symbol in both cases? We are, of course, able to recognise that the two sets of rules are intended to be rules for the same connective. But this depends on our previous understanding of the connectives, while the aim was to specify their meanings completely in terms of the rules governing them. It should not be down to our grasp of their meanings that we can recognise which rules belong to which connective, but solely down to the meaning-theoretical framework. Without addressing this question, bilateralists cannot claim that the meanings of the connectives of $\mathfrak{B}$ are determined completely by the rules of inference governing them, and should this be their objective, they have no right to use the same symbol in the two sets of rules governing a connective. 

Inversion principles that link the assertive and the rejective rules for a connective would answer the question raised in the previous paragraphs. There is, however, also another possible diagnosis of what has gone wrong in the four deductions, and this leads to a result of independent interest. In each of them, a complex formula is the conclusion of reductio and major premise of an elimination rule. Reductio provides grounds for the assertion and denial of formulas. These should be in harmony with the consequences of asserting and denying them as specified by the respective elimination rules for their main connective. This motivates the demand that formulas that are conclusion of reductio and major premise of an elimination rule should be removable from deductions. 

Furthermore, inferences by non-contradiction draw consequences from formulas which should be in harmony with the grounds for deriving them. Finally, reductio and non-contradiction should presumably be in harmony with each other, too, although this has nothing to do with the connectives, but rather with the formal framework of bilateral logic. 

These formulas also count as maximal in the normalisation theorem for deductions in $\mathfrak{B}$ that is proved in the next section.

\section{Normalisation for $\mathfrak{B}$} 
This section contains a proof of a normalisation theorem for deductions in $\mathfrak{B}$.\footnote{The reader is invited to compare it with St\aa lmarck's proof of normalisation for unilateral classical logic \citep{stalmarckclassical}. There is some resemblance, if $-$ is read as negation. However, as $\mathfrak{B}$ has a larger number of operational rules than St\aa lmarck's system, certain complications that arise in St\aa lmarck's proof do not arise here. In particular, there is no need to consider assumption contractions separately from reduction steps for maximal formulas. The larger number of rules also requires reduction steps for which there are no equivalents in St\aa lmarck's proof.} 

\begin{definition}
\normalfont The \emph{degree} of a signed formula $+ \ A$ or $- \ A$ is the number of connectives occurring in $A$.
\end{definition}

\noindent $\bot$ is not a signed formula and gets degree $0$. 

\begin{definition}
\normalfont A \emph{maximal signed formula} is an occurrence of a formula in a deduction that is one of the following: 

\noindent (a) conclusion of an introduction rule and major premise of an elimination rule;  

\noindent (b) conclusion of reductio and major premise of an elimination rule; 

\noindent (c) conclusion of reductio and premise of non-contradiction;

\noindent (d) conclusion of an introduction rule and premise of non-contradiction the other premise of which is also the conclusion of an introduction rule. 
\end{definition}

\noindent For brevity, I will mostly use `maximal formula' instead of `maximal signed formula'. 

To distinguish the four kinds of maximal formulas, I will call those of kind (a) \emph{maximal formulas with introduction and elimination rules} or \emph{i/e maximal formulas}; those of kind (b) \emph{maximal formulas with reductio and elimination rules} or \emph{r/e maximal formulas}; those of kind (c) \emph{maximal formulas with reductio and non-contradiction} or \emph{r/nc maximal formulas}; and those of kind (d) \emph{maximal formulas with introduction rules and non-contradiction} or \emph{i/nc maximal formulas}. 

Formulas of the third and fourth kind are clearly `maximal' in some sense, even though the philosophical reasons for requiring the removability of maximal formulas of the first (and perhaps the second) kind may not apply to them. They have been included here to ensure that deductions in normal form have the subformula property. For Rumfitt, i/nc maximal formulas do not arise, as he restricts non-contradiction to atomic premises. The reduction steps to remove r/e maximal formulas where the elimination rule is $+\lor E$ or $-\land E$ require the general version of non-contradiction. 

\begin{definition}[Segment, its Length and Degree, Maximal Segment]\ \\
\normalfont (a) A \emph{segment} is a sequence of two or more formula occurrences $C_1\ldots C_n$ in a deduction such that $C_1$ is not the conclusion of $+\lor E$ or $-\land E$, $C_n$ is not the minor premise of $+\lor E$ or $-\land E$, and for every $i<n$, $C_i$ is minor premise of $+\lor E$ or $-\land E$ and $C_{i+1}$ its conclusion. 

\noindent (b) The \emph{length} of a segment is the number of formula occurrences of which it consists, its \emph{degree} is their degree. 

\noindent (c) A segment is \emph{maximal} if and only if its last formula is major premise of an elimination rule or premise of non-contradiction.
\end{definition}

\noindent I will say that the formula occurrence $C_i$ is \emph{on} segment $C_1\ldots C_n$. A segment is above another one in a deduction if its last formula is above the other's first formula. I will speak of segments being the premises or conclusions of the rules of which their last or first formulas are premises or conclusions. 

Prawitz only counts a segment as maximal if it begins with the conclusion of an introduction rule \citep[49]{prawitznaturaldeduction}. The more general notion used here is also used by Troestra and Schwichtenberg in the proof of normalisation for intuitionist logic \citep[179]{troelstraschwichtenberg}. For philosophical reasons, the more general notion is called for, as it must be ensured that $+\lor E$ and $- \land E$ do not introduce grounds for the derivation of formulas that are not balanced by the elimination rules for their main operators. This is irrespective of how the first formula of the segment is derived.\footnote{See \citep[Ch 2]{kurbisproofandfalsity} for more on these philosophical reasons.} 

\begin{definition}
\normalfont A deduction is in \emph{normal form} if it contains neither maximal formulas nor maximal segments. 
\end{definition}

\noindent The reduction steps to be given next remove maximal formulas and maximal segments from deductions. Applying then in the systematic fashion specified in the proof of the normalisation theorem transforms any deduction into a deduction in normal form. $\leadsto$ indicates that the deduction to its left or above it is transformed into the deduction on its right or below it. I will call the deduction to which a reduction step is applied the \emph{original deduction} and the result of the application the \emph{reduced deduction}. A formula in square brackets between two deductions 

\begin{prooftree}
\AxiomC{$\Pi$}
\noLine
\UnaryInfC{$[A]$}
\noLine
\UnaryInfC{$\Sigma$}
\end{prooftree} 

\noindent means that the deduction on top is used to conclude all formulas in the assumption class $[A]$.\bigskip

\noindent (A) \emph{Permutative Reduction Steps for Maximal Segments} 

\noindent The lower application of the elimination rule or of non-contradiction is permuted upwards to conclude with a minor premise of $+\lor E$ or $-\land I$. Here are two examples, the others being similar. 

(1) The maximal segment consists of formula occurrences of the form $+\ C\lor D$, the last of which is concluded by $-\land E$:

\small
\begin{prooftree} 
\def\defaultHypSeparation{\hskip .1in}
\AxiomC{$\Xi_1$}
\noLine
\UnaryInfC{$- \ A\land B$} 
\AxiomC{$[- \ A]^i$}
\noLine
\UnaryInfC{$\Pi_1$}
\noLine
\UnaryInfC{$+\ C\lor D$} 
\AxiomC{$[- \ B]^j$}
\noLine
\UnaryInfC{$\Pi_2$}
\noLine
\UnaryInfC{$+\ C\lor D$}
\RightLabel{$_{i, j}$}
\TrinaryInfC{$+\ C\lor D$}
\AxiomC{$[+ \ C]^k$}
\noLine
\UnaryInfC{$\Sigma_1$}
\noLine
\UnaryInfC{$\phi$} 
\AxiomC{$[+ \ D]^l$}
\noLine
\UnaryInfC{$\Sigma_2$}
\noLine
\UnaryInfC{$\phi$}
\RightLabel{$_{k, l}$}
\TrinaryInfC{$\phi$}
\noLine
\UnaryInfC{$\Xi_2$}
\end{prooftree} 

\begin{center}
$\leadsto$
\end{center} 

\begin{prooftree} 
\def\defaultHypSeparation{\hskip .1in}
\AxiomC{$\Xi_1$}
\noLine
\UnaryInfC{$- \ A\land B$} 
\AxiomC{$[- \ A]^i$}
\noLine
\UnaryInfC{$\Pi_1$}
\noLine
\UnaryInfC{$+\ C\lor D$} 
\AxiomC{$[+ \ C]^{k_1}$}
\noLine
\UnaryInfC{$\Sigma_1$}
\noLine
\UnaryInfC{$\phi$} 
\AxiomC{$[+ \ D]^{l_1}$}
\noLine
\UnaryInfC{$\Sigma_2$}
\noLine
\UnaryInfC{$\phi$}
\RightLabel{$_{k_1, l_1}$}
\TrinaryInfC{$\phi$}
\AxiomC{$[- \ B]^j$}
\noLine
\UnaryInfC{$\Pi_2$}
\noLine
\UnaryInfC{$+\ C\lor D$}
\AxiomC{$[+ \ C]^{k_2}$}
\noLine
\UnaryInfC{$\Sigma_1$}
\noLine
\UnaryInfC{$\phi$} 
\AxiomC{$[+ \ D]^{l_2}$}
\noLine
\UnaryInfC{$\Sigma_2$}
\noLine
\UnaryInfC{$\phi$}
\RightLabel{$_{k_2, l_2}$}
\TrinaryInfC{$\phi$}
\RightLabel{$_{i, j}$}
\TrinaryInfC{$\phi$}
\noLine
\UnaryInfC{$\Xi_2$}
\end{prooftree} 

\normalsize
\noindent If some occurrence of $\phi$ forms part of a maximal segment in the original deduction, the permutative reduction step increases its length in the reduced deduction. In the proof of the normalisation theorem a strategy will be given to avoid increasing the length of a maximal segment of the same or higher degree than the one shortened or removed: in a nutshell, apply the reduction step to the rightmost segment of highest degree first. Furthermore, it needs to be ensured that the reduction step does not duplicate maximal formulas and segments of highest degree in $\Sigma_1$ and $\Sigma_2$: to do so it is applied to a topmost maximal segment of highest degree, one above which there is none other of highest degree. 

(2) The right premise of non-contradiction is conclusion of $-\land E$: 

\begin{prooftree}
\AxiomC{$\Sigma$}
\noLine
\UnaryInfC{$\alpha$}
\AxiomC{$\Xi_1$}
\noLine
\UnaryInfC{$- \ A\land B$} 
\AxiomC{$[- \ A]^i$}
\noLine
\UnaryInfC{$\Pi_1$}
\noLine
\UnaryInfC{$\alpha^\ast$} 
\AxiomC{$[- \ B]^j$}
\noLine
\UnaryInfC{$\Pi_2$}
\noLine
\UnaryInfC{$\alpha^\ast$}
\RightLabel{$_{i, j}$}
\TrinaryInfC{$\alpha^\ast$}
\BinaryInfC{$\bot$}
\noLine
\UnaryInfC{$\Xi_2$}
\end{prooftree} 

\begin{center} 
$\leadsto$
\end{center} 

\begin{prooftree}
\AxiomC{$\Xi_1$}
\noLine
\UnaryInfC{$- \ A\land B$} 
\AxiomC{$\Sigma$}
\noLine
\UnaryInfC{$\alpha$}
\AxiomC{$[- \ A]^i$}
\noLine
\UnaryInfC{$\Pi_1$}
\noLine
\UnaryInfC{$\alpha^\ast$} 
\BinaryInfC{$\bot$}
\AxiomC{$\Sigma$}
\noLine
\UnaryInfC{$\alpha$}
\AxiomC{$[- \ B]^j$}
\noLine
\UnaryInfC{$\Pi_2$}
\noLine
\UnaryInfC{$\alpha^\ast$}
\BinaryInfC{$\bot$}
\RightLabel{$_{i, j}$}
\TrinaryInfC{$\bot$}
\noLine
\UnaryInfC{$\Xi_2$}
\end{prooftree} 

\noindent The reduction step shortens the right segment, but if the left premise of non-contradiction is a maximal formula or the last formula of a segment, it duplicates it. As $\alpha$ and $\alpha^\ast$ have the same degree, it needs to be ensured that the step actually reduces the complexity of the deduction. So for the purpose of the proof of normalisation, the right premise of reductio will be counted as having a degree of one higher than the left premise, if both premises are maximal. This decides the question to which premise of reductio a reduction step is applied first in this and other cases.\bigskip

\noindent (B) \emph{Reduction Steps for Maximal Formulas} 

\noindent (a) \emph{Reduction Steps for Maximal Formulas with Introduction and Elimination Rules} 

\noindent These are not essentially different from those for intuitionist logic given by Prawitz, except that now a $+$ or $-$ is carried along in front of formulas, and there are additional reduction steps for the signed negations of formulas. The reduction steps for maximal formulas of the forms $+ \ A\lor B$ and $- \ A\land B$ are similar to those Prawitz gives for disjunctions, those for maximal formulas of the forms $- \ A \lor B$, $+ \ A\land B$ and $- \  A\supset B$ are similar to those Prawitz gives for conjunctions, those for maximal formulas of the form $+ \ A \supset B$ are similar to those Prawitz gives for implications, and the reduction steps for maximal formulas of the forms $+ \ \neg A$ and $- \ \neg A$ are evident enough. Applying such a reduction step may introduce new maximal formulas and segments into the reduced deduction, but they are of lower degree than the maximal formula removed from the original deduction. In cases of maximal formulas of the form $+ \ A\supset B$, $+\ A\lor B$ and $- \ A\land B$, the reduced deduction may contain multiple copies of subdeductions of the original deduction: to avoid multiplying maximal formulas or segments of the same or higher degree than the one removed, in the proof of the normalisation theorem the reduction steps are applied to maximal formulas of highest degree such that no maximal formulas or segments of highest degree stand above them or above the minor premises of the elimination rule of which they are the major premises.\bigskip 

\noindent (b) \emph{Reduction Steps for Maximal Formulas with Reductio and Elimination Rules} 

\noindent In the first three reduction steps below, if $+ \ A$ or $-\ A$ is major premise of an elimination rule or premise of non-contradiction in $\Sigma$, the reduction step introduces a new r/e or r/nc maximal formula of lower degree than the one removed, which presents no problem for the proof of normalisation. In the fourth case, a more difficult issue arises. 

(1) The r/e maximal formula has the form $+\ A\land B$: 

\begin{center}
\AxiomC{$[- \ A\land B]^i$}
\noLine
\UnaryInfC{$\Pi$}
\noLine
\UnaryInfC{$\bot$}
\RightLabel{$_i$}
\UnaryInfC{$+ \ A\land B$}
\UnaryInfC{$+ \ A$}
\noLine
\UnaryInfC{$\Sigma$}
\DisplayProof\qquad$\leadsto$\qquad
\AxiomC{$[- \ A]^i$}
\UnaryInfC{$[- \ A\land B]$}
\noLine
\UnaryInfC{$\Pi$}
\noLine
\UnaryInfC{$\bot$}
\RightLabel{$_i$}
\UnaryInfC{$+ \ A$}
\noLine
\UnaryInfC{$\Sigma$} 
\DisplayProof
\end{center} 

\noindent If any occurrences of $- \ A\land B$ in the assumption class $[-\ A\land B]^i$ of the original deduction are major premises of $- \land E$, then the reduction step introduces new i/e maximal formulas into the reduced deduction that have the same degree as the r/e maximal formula removed from the original deduction. Remove them as part of the present reduction step by applying the reduction step for i/e formulas of the form $-\ A\land B$ to each of them immediately after the transformation above: this creates at worst new maximal formulas of lower degree than the ones removed. Similarly if any occurrences of $- \ A\land B$ in the assumption class $[-\ A\land B]^i$ of the original deduction are premises of non-contradiction the other premise of which is also derived by an introduction rule: then new i/nc maximal formulas are introduced into the deduction, which are removed immediately after the transformation above as part of the step, and then, as the reduction procedures for such formulas to be given below show, at worst maximal formulas of lower degree arise. 

The case where $+\ B$ has been derived by $+\land E$ is similar, and so are the cases for r/e maximal formulas of the form $- \ A\lor B$.

(2) The r/e maximal formula has the form $- \ \neg A$: 

\begin{center} 
\AxiomC{$[+ \ \neg A]^i$}
\noLine
\UnaryInfC{$\Pi$}
\noLine
\UnaryInfC{$\bot$}
\RightLabel{$_i$}
\UnaryInfC{$- \ \neg A$}
\UnaryInfC{$+ \ A$}
\noLine
\UnaryInfC{$\Sigma$} 
\DisplayProof\qquad$\leadsto$\qquad
\AxiomC{$[- \ A]^i$}
\UnaryInfC{$[+ \ \neg A]$}
\noLine
\UnaryInfC{$\Pi$}
\noLine
\UnaryInfC{$\bot$}
\RightLabel{$_i$}
\UnaryInfC{$+ \ A$}
\noLine
\UnaryInfC{$\Sigma$}
\DisplayProof
\end{center} 

\noindent As in case (1), the reduction step may introduce new i/e or i/nc maximal formulas of the same degree as the r/e maximal formula removed, and this is dealt with in the same way: apply the relevant reduction steps immediately after the transformation above as part of the reduction step for r/e maximal formulas of the form $- \ \neg A$. 

The case for r/e maximal formulas of the form $+\ \neg A$ is similar. 

(3) There are three options for maximal formulas arising from reductio and elimination rules for implication: 

(i) The r/e maximal formula has the form $+ \ B\supset A$: 

\begin{center} 
\AxiomC{$[- \ B\supset A]^i$}
\noLine
\UnaryInfC{$\Pi$}
\noLine
\UnaryInfC{$\bot$}
\RightLabel{$_i$}
\UnaryInfC{$+ \ B\supset A$}
\AxiomC{$\Xi$}
\noLine
\UnaryInfC{$+ \ B$}
\BinaryInfC{$+\ A$}
\noLine
\UnaryInfC{$\Sigma$}
\DisplayProof\qquad$\leadsto$\qquad
\AxiomC{$\Xi$}
\noLine
\UnaryInfC{$+ \ B$}
\AxiomC{$[- \ A]^i$}
\BinaryInfC{$[- \ B\supset A]$}
\noLine
\UnaryInfC{$\Pi$}
\noLine
\UnaryInfC{$\bot$}
\RightLabel{$_i$}
\UnaryInfC{$+\ A$}
\noLine
\UnaryInfC{$\Sigma$}
\DisplayProof
\end{center} 

\noindent The reduction step may introduce new i/e or i/nc maximal formulas of the same degree as the r/e maximal formula removed, and this is dealt with as in previous cases. 

(ii) The r/e maximal formula has the form $- \ B\supset A$ and $-\ A$ is concluded: 

\begin{center} 
\AxiomC{$[+ \ B\supset A]^i$}
\noLine
\UnaryInfC{$\Pi$}
\noLine
\UnaryInfC{$\bot$}
\RightLabel{$_i$}
\UnaryInfC{$- \ B\supset A$}
\UnaryInfC{$-\ A$}
\noLine
\UnaryInfC{$\Sigma$}
\DisplayProof\qquad$\leadsto$\qquad
\AxiomC{$[+\ A]^i$}
\UnaryInfC{$[+ \ B\supset A]$}
\noLine
\UnaryInfC{$\Pi$}
\noLine
\UnaryInfC{$\bot$}
\RightLabel{$_i$}
\UnaryInfC{$-\ A$}
\noLine
\UnaryInfC{$\Sigma$}
\DisplayProof
\end{center} 

\noindent The reduction step may introduce new i/e or i/nc maximal formulas of the same degree as the r/e maximal formula removed, and this is dealt with as in previous cases. 

(iii) The r/e maximal formula has the form $- \ A\supset B$ and $+\ A$ is concluded: 

\begin{center} 
\AxiomC{$[+ \ A\supset B]^i$}
\noLine
\UnaryInfC{$\Pi$}
\noLine
\UnaryInfC{$\bot$}
\RightLabel{$_i$}
\UnaryInfC{$- \ A\supset B$}
\UnaryInfC{$+\ A$}
\noLine
\UnaryInfC{$\Sigma$}
\DisplayProof\qquad$\leadsto$\qquad
\AxiomC{$[+ \ A]^i$}
\AxiomC{$[- \ A]^{ii}$}
\BinaryInfC{$\bot$}
\UnaryInfC{$+ \ B$}
\RightLabel{$_i$}
\UnaryInfC{$[+ \ A\supset B]$}
\noLine
\UnaryInfC{$\Pi$}
\noLine
\UnaryInfC{$\bot$}
\RightLabel{$_{ii}$}
\UnaryInfC{$+\ A$}
\noLine
\UnaryInfC{$\Sigma$}
\DisplayProof
\end{center} 

\noindent The reduction step may introduce new i/e or i/nc maximal formulas of the same degree as the r/e maximal formula removed, and this is dealt with as in previous cases.\footnote{If non-contradiction is restricted to atomic premises, then the reduction step is incomplete: if $A$ is not atomic, the application of non-contradiction must be replaced by applications of non-contradiction to atomic subformulas of $A$. This, however, poses no difficulty, as $A$ is of lower degree than the r/e maximal formula removed.}

(4) The r/e maximal formula has  the form $+ \ A\lor  B$: 
{\small
\begin{center}
\def\defaultHypSeparation{\hskip .1in}
\AxiomC{$[- \ A\lor B]^i$}
\noLine
\UnaryInfC{$\Xi$}
\noLine
\UnaryInfC{$\bot$}
\RightLabel{$_i$}
\UnaryInfC{$+ \ A\lor B$}
\AxiomC{$[+ \ A]^{ii}$}
\noLine
\UnaryInfC{$\Pi_1$}
\noLine
\UnaryInfC{$\alpha$}
\AxiomC{$[+ \ B]^{iii}$}
\noLine
\UnaryInfC{$\Pi_2$}
\noLine
\UnaryInfC{$\alpha$}
\RightLabel{$_{ii, iii}$}
\TrinaryInfC{$\alpha$}
\noLine
\UnaryInfC{$\Sigma$}\DisplayProof\quad$\leadsto$\quad\def\defaultHypSeparation{\hskip .1in}\AxiomC{$[+ \ A]^i$}
\noLine
\UnaryInfC{$\Pi_1$}
\noLine
\UnaryInfC{$\alpha$}
\AxiomC{$[\alpha^\ast]^{iii}$}
\BinaryInfC{$\bot$}
\RightLabel{$_i$}
\UnaryInfC{$- \ A$}
\AxiomC{$[+ \ B]^{ii}$}
\noLine
\UnaryInfC{$\Pi_2$}
\noLine
\UnaryInfC{$\alpha$}
\AxiomC{$[\alpha^\ast]^{iii}$}
\BinaryInfC{$\bot$}
\RightLabel{$_{ii}$}
\UnaryInfC{$- \ B$}
\BinaryInfC{$[- \ A\lor B]$}
\noLine
\UnaryInfC{$\Xi$}
\noLine
\UnaryInfC{$\bot$}
\RightLabel{$_{iii}$}
\UnaryInfC{$\alpha$}
\noLine
\UnaryInfC{$\Sigma$}
\DisplayProof
\end{center} 
}
\noindent If $\alpha$ in the original deduction is $\bot$, non-contradiction is not applicable, but also not necessary: conclude $- \ A$ and $- \ B$ directly by reductio in the reduced deduction.  

The reduction step for r/e maximal formulas of the form $- \ A\land B$ is similar. 

$\Pi_1$ and $\Pi_2$ get multiplied as many times as there are assumptions in assumption class $[- \ A\lor B]^i$, so it must be ensured that when choosing a maximal formula to which to apply the reduction step, $\Pi_1$ and $\Pi_2$ contain no maximal formulas or segments of highest degree. The same strategy indicated for i/e maximal formulas works here: choose a maximal formula of highest degree such that no maximal formula or segment of highest degree stands above it or above the minor premises of the elimination rule of which it is the major premise. 

The reduction step may introduce new i/e or i/nc maximal formulas of the same degree as the r/e maximal formula removed, and this is dealt with as in previous cases. There are also three further cases to be considered. 

First, if $\alpha$ is major premise of an elimination rule in $\Sigma$, the reduction step may introduce an r/e maximal formula of unknown degree into the reduced deduction. In that case, however, $\alpha$ is the last formula of a maximal segment in the original deduction. To show that any deduction can be brought into one in normal form, the proof of the normalisation theorem describes a method that systematically removes all maximal formulas and segments from a deduction, beginning with those of highest degree: thus if the reduction step is applied as part of this process, $\alpha$ cannot be of higher degree than $+ \ A\lor B$. In the reduction steps for maximal segments and i/e maximal formulas it was noted that they are applied to maximal formulas of highest degree such that no maximal formulas of highest degree stand above them or the minor premises of the rule of which they are major premises. We need to ensure that in case the occurrence of $\alpha$ in $\Sigma$ is the last formula of a maximal segment of the same degree as $+\ A\lor B$ or forms part of such a maximal segment that continues in $\Sigma$, then the relevant permutative reduction step is applied to the segment first. The procedure indicated in the permutative reduction steps works here, too. If both have no maximal formulas or segments of highest degree above them, we apply the relevant reduction step to the rightmost one first, that is to one of which $\alpha$ forms part in this case. It'll be made more precise what `rightmost' means in the proof of the normalisation theorem.

Second, if $\alpha$ is the conclusion of reductio in $\Pi_1$ or $\Pi_2$, the reduction step may introduce an r/nc maximal formula of unknown degree into the reduced deduction. In that case, the application of non-contradiction in the reduced deduction is redundant and dropped from the reduction step. For example, suppose the last application of a rule in $\Pi_1$ is reductio. Then $\Pi_1$ has a subdeduction that derives $\bot$ from assumption classes $[+ \ A]^i$ and $[\alpha^\ast]$, so conclude $- \ A$ directly by reductio, discharging formulas in the assumption class $[+ \ A]^i$, and assign the formulas in the assumption class $[\alpha^\ast]$ in $\Pi_1$ to the new assumption class $iii$ and discharge them at the application of reductio that concludes with the $\alpha$ on top of $\Sigma$. Similarly if the last application of a rule in $\Pi_2$ is reductio, and if that is the last rule in both. 

Third, if $\alpha$ is the last formula of a segment in $\Pi_1$ or $\Pi_2$, then the reduction step introduces a new maximal segment into the deduction: remove it by permuting the application of non-contradiction upwards as described in the permutative reduction steps above as part of the reduction step. There remains one troublesome case to be taken care of: if the first formula of the segment is concluded by reductio in the original deduction, permuting non-contradiction upwards introduces an r/nc maximal formula of unknown degree into the reduced deduction. A version of the strategy of the previous paragraph works in this case, too. If the first formula of the segment is derived by reductio, we already have a subdeduction $\Pi_1'$ of $\Pi_1$ of $\bot$ from $[+ \ A]^i$ and $[\alpha^\ast]$ or a subdeduction $\Pi_2'$ of $\Pi_2$ of $\bot$ from $[+ \ B]^{ii}$ and $[\alpha^\ast]$. So conclude $- \ A$ or $- \ B$ directly by reductio without the redundant step of non-contradiction and assign the formulas in the assumption class $[\alpha^\ast]$ of $\Pi_1'$ or $\Pi_2'$ to assumption class $iii$, discharging them at the lower application of reductio marked in the reduction step. This leaves those assumptions in $\Pi_1'$ or $\Pi_2'$ undischarged that were discharged by applications of $+\lor E$ or $-\land E$ that gave rise to the segments in $\Pi_1$ or $\Pi_2$: so insert these applications before continuing with $\Sigma$, using the conclusion $\alpha$ of the lower application of reductio as the required minor premise. If $\alpha$ is on a segment in $\Sigma$, this increases its length. But notice that such a segment is either not maximal or of lower degree than the r/e maximal formula removed, by the choice of the strategy of choosing maximal segments or formulas in the proof of normalisation.\bigskip

\noindent (c) \emph{Reduction Steps for Maximal Formulas with Reductio and Non-Contradiction} 

\noindent There are two options to be considered. 

(1) The assumption discharged by the application of reductio is not premise of non-contradiction. I give as an example the case where the left premise of non-contradiction is a denial derived by reductio: 

\begin{center} 
\AxiomC{$\Sigma$}
\noLine
\UnaryInfC{$+ \ A$}
\AxiomC{$[+\ A]^i$}
\noLine
\UnaryInfC{$\Pi$}
\noLine
\UnaryInfC{$\bot$}
\RightLabel{$_i$}
\UnaryInfC{$- \ A$}
\BinaryInfC{$\bot$}
\noLine
\UnaryInfC{$\Xi$}
\DisplayProof\quad$\leadsto$\quad
\AxiomC{$\Sigma$}
\noLine
\UnaryInfC{$[+ \ A]$}
\noLine
\UnaryInfC{$\Pi$}
\noLine
\UnaryInfC{$\bot$}
\noLine
\UnaryInfC{$\Xi$}
\DisplayProof
\end{center} 

\noindent If any of the formulas in the assumption class $[+ \ A]^i$ is the major premise of an elimination rule in $\Pi$ and $+ \ A$ is the conclusion of an introduction rule or of reductio in $\Sigma$, then the reduction step introduces i/e or r/e maximal formulas into the reduced deduction. However, in this case both premises of non-contradiction are maximal, and the right one will be counted as one degree higher as the left one, and so the maximal formula created by the reduction step is of lower degree than the one removed. Similarly if $+ \ A$ is conclusion of $+\lor E$ or $-\land E$ in $\Sigma$ and any of the formulas in the assumption class $[+ \ A]^i$ is major premise of an elimination rule in $\Pi$: the new maximal segment is of degree one lower than the formula removed. 

If the situation is the mirror image of the one displayed and reductio concludes the left premise of non-contradiction, then the right premise is not conclusion of an elimination rule, as that one would be removed first. 

(2) The assumption discharged by reductio is premise of non-contradiction, say it is the left one: 

\begin{center} 
\AxiomC{$\Sigma$}
\noLine
\UnaryInfC{$+ \ A$}
\AxiomC{$[+\ A]^i$}
\AxiomC{$\Pi'$}
\noLine
\UnaryInfC{$- \ A$} 
\BinaryInfC{$\bot$} 
\noLine
\UnaryInfC{$\Pi''$}
\noLine
\UnaryInfC{$\bot$}
\RightLabel{$_i$}
\UnaryInfC{$- \ A$}

\BinaryInfC{$\bot$}
\noLine
\UnaryInfC{$\Xi$}
\DisplayProof\quad$\leadsto$\quad
\AxiomC{$\Sigma$}
\noLine
\UnaryInfC{$+ \ A$}
\AxiomC{$\Pi'$}
\noLine
\UnaryInfC{$- \ A$} 
\BinaryInfC{$\bot$}
\noLine
\UnaryInfC{$\Pi''$}
\noLine
\UnaryInfC{$\bot$}
\noLine
\UnaryInfC{$\Xi$}
\DisplayProof
\end{center} 

\noindent This reduction step does only what it is supposed to do: it removes one maximal formula and introduces no complications.\bigskip

\noindent (d) \emph{Reduction Steps for Maximal Formulas with Introduction Rules and Non-Contra\-diction} 

\noindent Two examples should suffice, the other cases being similar or obvious. 

(1) One premise is derived by $+ \land I$, the other by $-\land I$: 

\begin{center} 
\AxiomC{$\Pi_1$}
\noLine
\UnaryInfC{$+ \ A$} 
\AxiomC{$\Pi_2$}
\noLine
\UnaryInfC{$+ \ B$} 
\BinaryInfC{$+ \ A\land B$}
\AxiomC{$\Sigma$}
\noLine
\UnaryInfC{$- \ B$}
\UnaryInfC{$- \ A\land B$}
\BinaryInfC{$\bot$}
\noLine
\UnaryInfC{$\Xi$}
\DisplayProof\quad$\leadsto$\quad
\AxiomC{$\Pi_2$}
\noLine
\UnaryInfC{$+ \ B$} 
\AxiomC{$\Sigma$}
\noLine
\UnaryInfC{$- \ B$}
\BinaryInfC{$\bot$}
\noLine
\UnaryInfC{$\Xi$}
\DisplayProof
\end{center} 

(2) One premise is derived by $+\supset I$, the other by $-\supset I$: 

\begin{center} 
\AxiomC{$[+ \ A]^i$}
\noLine
\UnaryInfC{$\Pi$}
\noLine
\UnaryInfC{$+ \ B$}
\RightLabel{$_i$}
\UnaryInfC{$+ \ A\supset B$}
\AxiomC{$\Sigma_1$}
\noLine
\UnaryInfC{$+ \ A$}
\AxiomC{$\Sigma_2$}
\noLine
\UnaryInfC{$- \ B$}
\BinaryInfC{$- \ A \supset B$}
\BinaryInfC{$\bot$}
\noLine
\UnaryInfC{$\Xi$}
\DisplayProof\quad$\leadsto$\quad
\AxiomC{$\Sigma_1$}
\noLine
\UnaryInfC{$[+ \ A]$}
\noLine
\UnaryInfC{$\Pi$}
\noLine
\UnaryInfC{$+ \ B$}
\AxiomC{$\Sigma_2$}
\noLine
\UnaryInfC{$- \ B$}
\BinaryInfC{$\bot$}
\noLine
\UnaryInfC{$\Xi$}
\DisplayProof
\end{center} 

\noindent Applying the reduction steps may introduce new maximal formulas into the reduced deduction, but they are of lower degree than the maximal formulas removed from the original deduction. Choice of maximal formula to which to apply the step avoids duplicating maximal formulas of highest degree in $\Sigma$.\bigskip 

\noindent This completes the reduction steps for maximal formulas.\bigskip 

\noindent (C) \emph{Simplification Conversions} 

\noindent Applications of $+\lor E$ and $-\land E$ with empty assumption classes are redundant and may be removed from deductions.\bigskip 

\noindent This completes the description of the transformations of deductions applied in normalisation. 

The degree of a maximal formula or segment that is the right premise of reductio the left premise of which is also a maximal formula or segment is the degree of the formula (on the segment) plus $1$. For all others, it is the degree of the formula (on the segment). This also settles the question to which premise reduction steps for i/nc maximal formulas are applied, although this is of comparatively minor significance. 

\begin{definition}[Rank of a Deduction]
\normalfont The \emph{rank} of a deduction $\Pi$ is the pair $\langle d, l\rangle$ where $d$ is the highest degree of any maximal formula or segment in $\Pi$, and $l$ is the sum of the number of maximal formulas and the sum of the lengths of all maximal segments in $\Pi$. If there are no maximal formulas or segments in $\Pi$, its rank is $0$.
\end{definition} 

\noindent Ranks are ordered lexicographically: $\langle d, l\rangle< \langle d', l'\rangle$ iff either $d<d'$ or $d=d'$ and $l<l'$. 

As we have been rather explicit about the considerations necessary to ensure that the complexity of a deduction is decreased in applying the reduction steps, the proof of normalisation itself can thankfully be brief. All that remains is to explicate the notion of a `rightmost' maximal formula or segment. Here we follow Prawitz \citep[50]{prawitznaturaldeduction}.  

\begin{theorem}
Any deduction $\Pi$ of $\alpha$ from $\Gamma$ in $\mathfrak{B}$ can be brought into a deduction in normal form of $\alpha$ from some of $\Gamma$.
\end{theorem} 

\noindent \emph{Proof.} By induction over the rank of deductions and applying the reduction steps. Take a maximal formula or maximal segment of highest degree such that (i) no maximal formula or segment of highest degree stands above it in the deduction, (ii) no maximal formula or segment of highest degree stands above a minor premise of the elimination rule of which the maximal formula or segment is the major premises, and (iii) no maximal segment of highest degree contains a formula that is minor premise of the elimination rule of which the maximal formula or maximal segment is the major premise. This reduces the rank of the deduction. \emph{Q.e.d.}

\section{Philosophical Assessment} 
There are at least two reasons why not everyone will be satisfied that the proof of section 5 solves the philosophical problems of section 4:\bigskip

\noindent (1) It appeals to non-contradiction in its general form. 

\noindent (2) The definition of `maximal signed formula' merely changes the topic.\bigskip 

\noindent Let's look at each charge in turn. 

In reduction step (B.b.4), the one for r/e maximal formulas of the form $+ \ A\lor B$ and $- \ A\land B$, non-contradiction is applied to arbitrary formulas $\alpha$. According to Rumfitt, if $\alpha$ is not atomic, the inference from $\alpha$ and $\alpha^\ast$ to $\bot$ needs to be replaced by applications of non-contradiction to atomic subformulas of $\alpha$. The difficulty is that this may introduce new maximal formulas of unknown degree into the deduction. Consider the construction that shows how to replace premises of the form $C\lor D$ by $C$ and $D$: 

\begin{center} 
\AxiomC{$+ \ C\lor D$}
\AxiomC{$- \ C\lor D$}
\BinaryInfC{$\bot$}
\DisplayProof\bigskip

$\leadsto$\bigskip

\AxiomC{$+\ C\lor D$}
\AxiomC{}
\RightLabel{$_i$}
\UnaryInfC{$+ \ C$}
\AxiomC{$- \ C\lor D$}
\UnaryInfC{$- \ C$}
\BinaryInfC{$\bot$}
\AxiomC{}
\RightLabel{$_i$}
\UnaryInfC{$+ \ D$}
\AxiomC{$- \ C\lor D$}
\UnaryInfC{$- \ D$}
\BinaryInfC{$\bot$}
\RightLabel{$_i$}
\TrinaryInfC{$\bot$}
\DisplayProof
\end{center} 

\noindent Suppose in the original deduction displayed in the reduction step (B.b.4), $\alpha$ is a disjunction on a segment that is the conclusion of $+ \lor I$ or $-\land I$. If this segment is major premise of $+\lor E$ or $-\land E$ or non-contradiction, all is fine: either $\alpha$ has lower degree than $- \ A\lor B$ or its segment is removed first. If it is not, however, then the procedure for removing complex premises of non-contradiction introduces maximal formulas of unknown degree into the reduced deduction. 

The bilateralist who insists on restricting non-contradiction to atomic premises requires a different proof of normalisation from the one given here. Alternatively, the bilateralist could treat $\land$ and $\lor$ as defined in terms of $\supset$ and $\neg$. One might also wonder whether the restriction of non-contradiction to atomic premises is an essential element of bilateralism. It is according to Rumfitt, but the current considerations may constitute a recommendation to drop it. 

Another option that solves the problems of section 4 would be to restrict reductio to atomic conclusions. Ferreira observes that once non-contradiction is restricted to atomic premises, there may be no good reason not to restrict reductio correspondingly \citep{ferreira}. Rumfitt's reasons for restricting non-contradiction seem to apply just as well to reductio. Reductio is a rule of the same kind as non-contradiction, a structural rule concerning the co-ordination of assertion and denial. 

Ferreira shows, however, that the resulting logic is not classical and contains neither $+\ A\lor \neg A$ nor $- \ A\land \neg A$ as theorems. This may not be so much a defect of bilateralism, as rather the surprising or interesting result that the correct logic of bilateralism is not classical logic, but a constructive logic with strong negation. This is the position for which Wansing argues \citep{wansingmoregeneral}. The current considerations may add support to this line of thought. It certainly has something to be said for it. It was noted by Gibbard that dropping reductio and non-contradiction altogether from $\mathfrak{B}$ gives a constructive logic with strong negation \citep{gibbard}. Reading $-$ as $\neg$ and ignoring $+$, it is Nelson's logic of constructible falsity, also discussed by Prawitz \citep[96f]{prawitznaturaldeduction}. While Wansing's logic adds further connectives, which require additional reduction steps, the proof of section 5 also gives normalisation theorems for logics arising from $\mathfrak{B}$ by dropping non-contradiction and reductio or restricting both to atomic formulas. 

In as much as bilateralism was supposed to justify classical logic, however, this line of argument is problematic. Much of the motivation for bilateralism is to overcome Dummettian objections to classical logic, in particular that the rules for classical negation are not stable. Many bilateralists will therefore prefer a different route to excluding \emph{honk} and \emph{conk}. 

Now for changing the subject. The requirement that r/e maximal formulas be removable from deductions is rather different from the similar requirement on i/e maximal formulas. The latter provides a formal criterion for fulfilment of Prawitz's inversion principle. Stability is a relation between the operational rules for a connective, its introduction and elimination rules. The unilateral approach locates any defects in rules for connectives in the operational rules governing them. The notion of a maximal signed formula incorporates a relation between one rule and all elimination rules. That one rule is a structural rule, concerning the formal framework of bilateralism, and so the notion of a maximal signed formula incorporates aspects of a rather different kind than those on which proof-theoretic semantics was originally built.  

This objection does, I think, show something, but not that something is wrong with the present notion of a maximal signed formula. It rather exhibits a shortcoming of bilateralism. There must obtain some balance in the inferential powers of reductio and the other rules. If the rather obvious way of capturing that balance employed here is objectionable, so much the worse for bilateralism. 

Where I would agree is that the solution proposed here does not really go to the heart of the matter of what is wrong with \emph{conk} and \emph{honk}. The problem with \emph{tonk} lies in the mismatch of its introduction and elimination rules. One would expect a comparable diagnosis of the problem with \emph{conk} and \emph{honk} from bilateralism: it lies in a mismatch of their assertive and their rejective rules. Locating the problem with \emph{conk} and \emph{honk} in reductio is not to the point. One should expect bilateral inversion principles that provide a general basis on which to diagnose mismatch of operational rules, just as the inversion principles in the unilateral context do, where these cut across the divide of assertive and rejective rules.

\section{Conclusion} 
The most promising solution to the problem of section 4 would be to formulate a bilateral notion of stability that incorporates bilateral inversion principles and a notion harmony between the assertive and the rejective rules of the connectives. 

One proposal of how to do this has been formulated by Francez \citep{francezbilateralharmony}. His notion of \emph{vertical harmony} holds between assertive introduction and elimination rules and rejective introduction and elimination rules, while \emph{horizontal harmony} holds between assertive and rejective introduction rules. Francez modifies horizontal harmony in a slightly later paper, where it is also noted that it provides a notion of harmony between rejective and assertive elimination rules \citep{francezgabbayresponse}. Another proposal is by the present author \citep{kurbisbilatinv}.\footnote{Both proposals allow the bilateralist to rule out the bilateral intuitionist logic of \citep{kurbisrumfitt}, the rules of which, it must be admitted, are not as nicely symmetrical as those of $\mathfrak{B}$.} 

There are, however, reasons to believe that adopting a bilateral notion of stability would be counterproductive for the bilateralist. 

In the unilateral framework, there are two aspects of the use, and thus meaning, of the connectives in deductive arguments: one is captured by the introduction rules and the other by the elimination rules for a connective. These aspects must be in harmony, or more precisely stable, and satisfy the inversion principles. Following Gentzen, the introduction rules for the connectives define their meanings, and the elimination rules are consequences thereof. Following Dummett and Prawitz, they are consequences in the sense that they are determined from the introduction rules by the inversion principles. As stability is a requirement on rules that are to define the meanings of the connectives completely, the process could be reversed and the elimination rules taken as prior and the introduction rules determined from them. 

Transpose this to the bilateral case. The motivating thesis of bilateralism is that the meanings of the expressions of a language are determined by the conditions of the correct assertibility and the correct deniability of sentences of which they form part. The bilateralist agrees that stability must obtain between the introduction and elimination rules for the connectives. Let's follow Gentzen again and pick the introduction rules as those that define the meaning of a connective, while its elimination rules are consequences of them by the bilateral notion of stability. \emph{honk} and \emph{conk} show that we cannot simply lay down assertive and rejective introduction rules for a connective. They, too, must be balanced by the bilateral notion of stability. But this means that only one kind of introduction rules defines the meaning of the connective, and the other is a consequence by bilateral stability. 

In the absence of a principled way of deciding between the two kinds of introduction rules, we might as well pick the assertive introduction rules as defining the meanings of the connectives, all others being determined from them by bilateral stability. And now the situation looks awkward for the bilateralist. The bilateralist claims that the meanings of the connectives are defined by the assertive and rejective rules governing them. A closer look into the matter reveals that they are defined by the assertive introduction rules. That is exactly the thesis of the unilateralist. All introduction rules of unilateral logic are assertive. 

Nothing hangs on the choice of assertive introduction rules as defining meaning. To rule out \emph{conk} and \emph{honk}, and to emulate the notion of stability of the unilateral approach, the bilateralist needs inversion principles that determine the three other sets of rules for a connective from any given one. Still, it is only one aspect of the use of the connective that defines its meaning, the others being consequences by stability, not two of them, as claimed by the bilateralist. It is not the assertive rules in tandem with rejective rules that determine the meaning of a connective, but only one half of one of those two aspects -- either the assertive introduction rules, or the assertive elimination rules, or the rejective introduction rules, or the rejective elimination rules -- the rest being determined by bilateral stability. Thus it looks as if adopting a bilateral notion of stability means that the characteristically bilateral thesis on how meanings are determined is effectively abandoned, and bilateralism collapses into a form of unilateralism. 

This looks like a dilemma for bilateralists. Formulate a bilateral notion of stability, or else face \emph{conk} and \emph{honk}. But if you do the former, face giving up bilateralism.

\bigskip

\setlength{\bibsep}{0pt}
\bibliographystyle{chicago}
\bibliography{KurbisBilateralTonk}

\begin{thebibliography}{}

\bibitem[\protect\citeauthoryear{Dummett}{Dummett}{1993a}]{dummettLBM}
Dummett, M. (1993a).
\newblock {\em The Logical Basis of Metaphysics}.
\newblock Cambridge, Mass.: Harvard University Press.

\bibitem[\protect\citeauthoryear{Dummett}{Dummett}{1993b}]{dummettmeaningI}
Dummett, M. (1993b).
\newblock {What is a Theory of Meaning? (I)}.
\newblock In {\em The Seas of Language}, pp.\  1--33. Oxford: Clarendon.

\bibitem[\protect\citeauthoryear{Dummett}{Dummett}{1993c}]{dummettmeaningII}
Dummett, M. (1993c).
\newblock {What is a Theory of Meaning? (II)}.
\newblock In {\em The Seas of Language}, pp.\  34--93. Oxford: Clarendon.

\bibitem[\protect\citeauthoryear{Ferreira}{Ferreira}{2008}]{ferreira}
Ferreira, F. (2008).
\newblock The co-ordination principles: A problem for bilateralism.
\newblock {\em Mind\/}~{\em 117\/}(468), 1051--1057.

\bibitem[\protect\citeauthoryear{Francez}{Francez}{2014}]{francezbilateralharmony}
Francez, N. (2014).
\newblock Bilateralism in proof-theoretic semantics.
\newblock {\em Journal of Philosophical Logic\/}~{\em 43\/}(2/3), 239--259.

\bibitem[\protect\citeauthoryear{Francez}{Francez}{2018}]{francezgabbayresponse}
Francez, N. (2018).
\newblock Bilateralism does provide a proof theoretic treatment of classical
  logic (for non-technical reasons).
\newblock {\em Journal of Applied Logics\/}~{\em 5\/}(8), 1653--1662.

\bibitem[\protect\citeauthoryear{Gabbay}{Gabbay}{2017}]{gabbaybilateralism}
Gabbay, M. (2017).
\newblock Bilateralism does not provide a proof theoretic treatment of
  classical logic (for technical reasons).
\newblock {\em Journal of Applied Logic\/}~{\em 25\/}(S), 108--122.

\bibitem[\protect\citeauthoryear{Gentzen}{Gentzen}{1934}]{gentzenuntersuchungen}
Gentzen, G. (1934).
\newblock {Untersuchungen \"uber das logische Schlie\ss en. I}.
\newblock {\em Mathematische Zeitschrift\/}~{\em 39\/}(2), 176--210.

\bibitem[\protect\citeauthoryear{Gibbard}{Gibbard}{2002}]{gibbard}
Gibbard, P. (2002).
\newblock Price and {Rumfitt} on rejective negation and classical logic.
\newblock {\em Mind\/}~{\em 111\/}(442), 297--303.

\bibitem[\protect\citeauthoryear{Humberstone}{Humberstone}{2000}]{humberstonerejection}
Humberstone, L. (2000).
\newblock The revival of rejective negation.
\newblock {\em Journal of Philosophical Logic\/}~{\em 29\/}(4), 331--381.

\bibitem[\protect\citeauthoryear{K\"urbis}{K\"urbis}{2015}]{kurbisnegation}
K\"urbis, N. (2015).
\newblock What is wrong with classical negation?
\newblock {\em Grazer Philosophische Studien\/}~{\em 92\/}(1), 51--85.

\bibitem[\protect\citeauthoryear{K\"urbis}{K\"urbis}{2016}]{kurbisrumfitt}
K\"urbis, N. (2016).
\newblock Some comments on {Ian Rumfitt}'s bilateralism.
\newblock {\em Journal of Philosophical Logic\/}~{\em 45\/}(6), 623--644.

\bibitem[\protect\citeauthoryear{K\"urbis}{K\"urbis}{2017}]{kurbisbilateraldetours}
K\"urbis, N. (2017).
\newblock Bilateralist detours: From intuitionist to classical logic and back.
\newblock {\em Logique et Analyse\/}~{\em 60\/}(239), 301--316.

\bibitem[\protect\citeauthoryear{K\"urbis}{K\"urbis}{2019a}]{kurbisbilatinv}
K\"urbis, N. (2019a).
\newblock Bilateral inversion principles.
\newblock \emph{To be published}.

\bibitem[\protect\citeauthoryear{K\"urbis}{K\"urbis}{2019b}]{kurbisproofandfalsity}
K\"urbis, N. (2019b).
\newblock {\em Proof and Falsity. A Logical Investigation}.
\newblock Cambridge University Press.

\bibitem[\protect\citeauthoryear{Milne}{Milne}{2015}]{milneinversion}
Milne, P. (2015).
\newblock Inversion principles and introduction rules.
\newblock In H.~Wansing (Ed.), {\em Dag Prawitz on Proofs and Meaning}, pp.\
  189--224. Cham, Heidelberg, New York, Dordrecht, London: Springer.

\bibitem[\protect\citeauthoryear{Moriconi and Tesconi}{Moriconi and
  Tesconi}{2008}]{moriconitesconi}
Moriconi, E. and L.~Tesconi (2008).
\newblock On inversion principles.
\newblock {\em History and Philosophy of Logic\/}~{\em 29\/}(2), 103--113.

\bibitem[\protect\citeauthoryear{Negri and von Plato}{Negri and von
  Plato}{2001}]{negriplatostructural}
Negri, S. and J.~von Plato (2001).
\newblock {\em Structural Proof-Theory}.
\newblock Cambridge University Press.

\bibitem[\protect\citeauthoryear{Prawitz}{Prawitz}{1965}]{prawitznaturaldeduction}
Prawitz, D. (1965).
\newblock {\em Natural Deduction}.
\newblock Stockholm, G\"oteborg, Uppsala: Almqvist and Wiksell.

\bibitem[\protect\citeauthoryear{Price}{Price}{1983}]{pricesense}
Price, H. (1983).
\newblock Sense, assertion, {Dummett} and denial.
\newblock {\em Mind\/}~{\em 92\/}(366), 161--173.

\bibitem[\protect\citeauthoryear{Prior}{Prior}{1961}]{priorrunabout}
Prior, A. (1961).
\newblock The runabout inference ticket.
\newblock {\em Analysis\/}~{\em 21\/}(2), 38--39.

\bibitem[\protect\citeauthoryear{Rumfitt}{Rumfitt}{2000}]{rumfittyesno}
Rumfitt, I. (2000).
\newblock {``Yes'' and ``No''}.
\newblock {\em Mind\/}~{\em 109\/}(436), 781--823.

\bibitem[\protect\citeauthoryear{Smiley}{Smiley}{1996}]{smileyrejection}
Smiley, T. (1996).
\newblock Rejection.
\newblock {\em Analysis\/}~{\em 56\/}(1), 1--9.

\bibitem[\protect\citeauthoryear{{St\aa lmarck}}{{St\aa
  lmarck}}{1991}]{stalmarckclassical}
{St\aa lmarck}, G. (1991).
\newblock Normalisation theorems for full first order classical natural
  deduction.
\newblock {\em Journal of Symbolic Logic\/}~{\em 52\/}(2), 129--149.

\bibitem[\protect\citeauthoryear{Troestra and Schwichtenberg}{Troestra and
  Schwichtenberg}{2000}]{troelstraschwichtenberg}
Troestra, A. and H.~Schwichtenberg (2000).
\newblock {\em Basic Proof Theory\/} (2 ed.).
\newblock Cambridge University Press.

\bibitem[\protect\citeauthoryear{Wansing}{Wansing}{2017}]{wansingmoregeneral}
Wansing, H. (2017).
\newblock A more general general proof theory.
\newblock {\em Journal of Applied Logic\/}~{\em 25\/}(S), 23--46.

\end{thebibliography}


\begin{thebibliography}{}

\bibitem[\protect\citeauthoryear{K\"urbis}{K\"urbis}{2021}]{kurbisbilateralnormalisation}
K\"urbis, N. (2021).
\newblock Normalisation for bilateral classical logic with some philosophical
  remarks.
\newblock {\em The Journal of Applied Logics\/}~{\em 8\/}(2), 531--556.

\bibitem[\protect\citeauthoryear{{St\aa lmarck}}{{St\aa
  lmarck}}{1991}]{stalmarckclassical}
{St\aa lmarck}, G. (1991).
\newblock Normalisation theorems for full first order classical natural
  deduction.
\newblock {\em Journal of Symbolic Logic\/}~{\em 52\/}(2), 129--149.

\end{thebibliography}

\end{document}